\documentclass[aps,showpacs,preprintnumbers,amsmath,amssymb,nofootinbib]{revtex4}
\usepackage{bm}

\begin{document}

\title{Induced inflation from a 5D purely kinetic scalar field formalism on warped product spaces}

\author{Jos\'e Edgar Madriz Aguilar\footnote{E-mail address: jemadriz@fisica.ufpb.br}}
\affiliation{Departamento de F\'{\i}sica, Universidade Federal da
Para\'{\i}ba. C.P. 5008-CEP: 58059-970, \\ Jo\~{a}o Pessoa, PB
58059-970 Brazil.}

\vskip .5cm
\begin{abstract}
Considering a separable and purely kinetic 5D scalar field we investigate the induction of 4D scalar potentials on a 4D constant 
foliation on the class of 5D warped product space-times. We obtain a quantum confinement of the inflaton modes given naturally 
from the model for at least a class of warping factors.  We can recover a 4D inflationary scenario where the inflationary 
potential is geometrically induced from 5D and the effective equation of state in 4D that includes the effect of the inflaton 
field and the induced matter is $P_{eff}\simeq -\rho _{eff}$.
\end{abstract}


\maketitle

\section{Introduction}

Warped geometries are the natural geometric background of Randall-Sundrum scenarios \cite{RS1}. In the braneworld scenarios this 
kind of geometries has been subject of great interest. Mathematically, given $(M^{a},\gamma)$ and $(M^{b},h)$ two Riemannian (or 
pseudo-Riemannian) manifolds of dimensions $a$ and $b$, and metrics $\gamma$ and $h$, respectively, it is possible to construct a 
new Riemannian (or pseudo-Riemannian) manifold by $(M=M^{a}\times M^{b},g)$ where by definition $g=e^{2A}\gamma\otimes h$, being 
$A:M^{b}\rightarrow\Re$ a given smooth function known as the warping factor \cite{CRomero}. An interesting property of this kind 
of geometries is
the natural splitting that naturally occurs between the dynamics in the 4D space-time and the dynamics in the extra dimensions. In 
the particular case of 5D spaces we can separate the motion in the fifth coordinate from the one in the remaining 4D usual 
space-time dimensions.  This feature could be of great ``wealth" in some of the new treatments of inflation in theories in more 
than four dimensions.\\

Some of the most popular theories in higher dimensions are the Kaluza-Klein theory \cite{KK} where the fifth dimension is 
compactified, the induced matter theory or in more general terms the noncompactified (n+1)-general relativity \cite{Wes} and the 
braneworld scenarios inspired in a string theory known as the M-theory \cite{MT}. The universe in braneworld cosmologies is 
considered as a brane embedded in a higher dimensional space-time called the bulk. In these models only gravity and some exotic 
matter like dilatonic scalar fields can propagate through the bulk, while our observable universe is confined to a particular 
brane hypersurface \cite{Brax}. Several mechanisms of confinement of matter to a special 3+1-hypersurface have been implemented. 
The use of a confining potential is one of these mechanisms \cite{Conf}. However, this continues being a central issue not only in 
braneworld scenarios but also in the induced matter theory.\\

New approaches to inflationary cosmology based on a scalar field formalism have been proposed in the context of higher dimensional 
theories of gravity. In particular using ideas of the induced matter theory a new formalism for describing inflation \cite{n1}, 
scalar metric fluctuations and gravitational waves \cite{n2}, has been recently introduced. Some other topics as for example  
effects of a decaying cosmological parameter in this recent 5D framework were also studied \cite{n3}. The basic idea of this 
formalism is the existence of a 5D space-time equipped with a purely kinetic scalar field $\varphi$, in which our observable 4D 
universe can be confined to a particular hypersurface. The induced 4D inflationary dynamics is described in terms of the induced 
4D scalar field $\varphi (x,\psi _{0})$, being $\psi _0$ the value of the fifth coordinate specified by a given foliation of the 
5D space-time, where the particular choice of the hypersurface $\psi =H^{-1}$, with $H$ the Hubble parameter, is crucial on its 
physical predictions.\\

In this letter we develop a new approach in a more general manner. We start considering a purely kinetic 5D scalar field on a 
warped product geometrical background. Assuming the separability of the scalar field and that the 5D space-time can be foliated, 
we present a general mechanism for inducing a scalar potential of a true 4D scalar field $\phi(x)$ instead of a potential for a 4D 
effective scalar field $\varphi (x,\psi _0)$. As an application of the formalism we study de-Sitter inflationary cosmology 
recovering the usual 4D formalism but with a geometrically induced 4D scalar potential $V(\phi)$. As we will see, the quantum 
confinement of the inflaton modes is obtained naturally from the model and physical predictions can be made independent of the 
hypersurface chosen.

\section{General formalism}

We consider a 5D space-time $(M,g_{AB})$, where $g_{AB}$ is a 5D metric which in the local coordinates $( y^{A})$ is expressed by 
the line element $dS^{2}=g_{AB}dy^{A}dy^{B}$. The action that describes a purely kinetic 5D scalar field $\varphi (y^{A})$ 
minimally coupled to gravity can be written as
\begin{equation}\label{a1}
^{(5)}{\cal S}=\int d^{5}y \sqrt{|g|}\left[\frac{\,^{(5)}{\cal R}}{2\kappa^{2}}+\,^{(5)}{\cal L}(\varphi,\varphi _{,A})\right],
\end{equation}
where $^{(5)}{\cal R}$ is the scalar curvature, $\kappa$ defines the 5D gravitational coupling and the lagrangian density for the 
scalar field is
\begin{equation}\label{aa1}
^{(5)}L(\varphi,\varphi _{,A})=\sqrt{|g|}\,\,^{(5)}{\cal L}(\varphi ,\varphi _{,A})=\sqrt{|g|}\,\left[\frac{1}{2}g^{AB}\varphi 
_{,A}\varphi _{,B}\right].
\end{equation}
The variation of the action with respect to the metric and the scalar field, respectively, gives the 5D field equations
\begin{eqnarray}\label{a2}
G_{AB}&=&\kappa T_{AB},\\
\label{a3}
^{(5)}\Box \varphi &\equiv& \frac{1}{\sqrt{|g|}}\frac{\partial}{\partial y^{A}}\left[\sqrt{|g|}g^{AB}\varphi _{,B}\right]=0,
\end{eqnarray}
being $G_{AB}=R_{AB}-(1/2)\,^{(5)}{\cal R}g_{AB}$ the Einstein tensor in 5D and $T_{AB}$ the energy momentum tensor for a free 
scalar field specified by
\begin{equation}\label{a4}
T_{AB}=\varphi _{,A}\varphi _{,B}-\frac{1}{2}g_{AB}\varphi ^{,C}\varphi _{,C}.
\end{equation}
In the local coordinates $(y^{A})=(x^{\mu},\psi)$, we consider the class of warped geometries given by the line element
\begin{equation}\label{a5}
dS^{2}=e^{2A(\psi)}h_{\mu\nu}(x)dx^{\mu}dx^{\nu}-d\psi^{2},
\end{equation}
where $h_{\mu\nu}$ is a 4D metric with determinant $det(h_{\mu\nu})=h$.
On this metric background equation (\ref{a3}) reads
\begin{equation}\label{a6}
^{(4)}\Box \varphi -e^{-2A(\psi)}\frac{\partial}{\partial \psi}\left[e^{4A(\psi)}\frac{\partial\varphi}{\partial\psi}\right]=0,
\end{equation}
being $^{(4)}\Box\varphi=(1/\sqrt{-h})(\partial/\partial x^{\mu})(\sqrt{-h}\, h^{\mu\nu}\varphi _{,\nu})$.
Assuming a separable scalar field $\varphi(x,\psi)=\phi(x)Q(\psi)$, the expression (\ref{a6}) yields the system
\begin{eqnarray}\label{a7}
&& ^{(4)}\Box \phi(x)-\alpha\phi(x)=0,\\
\label{a8}
&&\frac{d^{2}Q}{d\psi^2}+4\frac{dA(\psi)}{d\psi}\frac{dQ}{d\psi}-\alpha e^{-2A(\psi)}Q=0,
\end{eqnarray}
with $\alpha$ a separation constant. This is a system of ordinary differential equations of second order which in principle, for a 
given warping factor $A(\psi)$, can be solved. \\

We can write equation (\ref{aa1}) as
\begin{equation}\label{a9}
^{(5)}L(\varphi,\varphi _{,A})=e^{2A(\psi)}Q^{2}(\psi)\sqrt{-h}\,\left[\frac{1}{2}h^{\mu\nu}\phi _{,\mu}\phi 
_{,\nu}-\frac{1}{2}e^{2A(\psi)}\left(\frac{\overset{\star}{Q}}{Q}\right)^{2}\phi^{2}(x)\right],
\end{equation}
where the star $(\star)$ denotes $\partial /\partial\psi$. Now assuming that the 5D space-time can be foliated by a family of 
hypersurfaces defined by $\Sigma _{0}:\psi=\psi _{0}$, being $\psi _{0}$ a constant, equation (\ref{a9}) evaluated on one of these 
hypersurfaces becomes
\begin{equation}\label{a10}
^{(4)}L(\phi,\phi _{,\mu})\equiv \,^{(5)}L_{\Sigma _0}(\varphi,\varphi _{,A})=e^{2A(\psi _0)}Q^{2}(\psi 
_0)\sqrt{-h}\left[\frac{1}{2}\,h^{\mu\nu}\phi _{,\mu}\phi _{,\nu}-V_{ind}(\phi)\right],
\end{equation}
where $V_{ind}(\phi)$ is the induced 4D scalar potential given by
\begin{equation}\label{a11}
V_{ind}(\phi)=\frac{1}{2}\,e^{2A(\psi _0)}\left.\left(\frac{\overset{\star}{Q}}{Q}\right)^{2}\right|_{\psi _0}\phi ^{2}(x),
\end{equation}
which is different for every warping factor. Clearly the induction of the potential depends strongly of the separability condition 
of $\varphi$ and of the splitting property of the warped product metrics. Hence the action (\ref{a1}) evaluated on $\Sigma _{0}$ 
reads
\begin{equation}\label{a13}
^{(4)}{\cal S}=\int d^{4}x\,\sqrt{-h}\left[\frac{^{(4)}{\cal R}}{16\pi G}+{\cal L}_{IM}+\frac{1}{2}h^{\mu\nu}\phi _{,\mu}\phi 
_{,\nu}-V_{ind}(\phi)\right],
\end{equation}
where we have made the identification  $\kappa ^{2}Q^{2}(\psi _0)=8\pi G$ and ${\cal L}_{IM}$ is a 4D induced matter lagrangian 
density coming from the part of $^{(5)}{\cal R}$ that depends of the fifth coordinate $\psi$ \cite{Wes}. The action (\ref{a13}) is 
interpreted as a 4D induced action, which is constructed on $\Sigma _0$ by means of the 4D induced lagrangian (\ref{a10}). This 
action besides of induce matter in a geometrical manner, allows to describe the dynamics of the 4D scalar field $\phi(x)$ 
minimally coupled to gravity with a geometrically induced scalar potential $V_{ind}(\phi)$. The fact of having a geometrical 
mechanism of deriving a scalar potential avoiding its introduction as ``by hand" is an attractive feature of this approach. 
Moreover, as we shall see in section IV, the fact of having a scalar potential induced geometrically bears to have a physical 
scalar field mass parameter also geometrical in origin and this is a valuable characteristic specially on inflationary frameworks. 
Compared with standard 4D  inflationary scenarios this feature represents an advantage in the sense that for example it can be 
possible to treat a DeSitter inflation without the introduction {\it a priori } of a little mass parameter for the inflaton field. 
This is basically because in this kind of formalism like the present one it is possible to treat a DeSitter expansion without a 
constant scalar potential by simply choosing a foliation $\psi =\psi _{0}=H_{0}^{-1}$, being $H_{0}$ the constant Hubble parameter 
\cite{ql}.\\

The field equations according to (\ref{a13}) are then
\begin{equation}\label{pr1}
^{(4)}G_{\mu\nu}=8\pi G\,^{(4)}T^{(\phi)}_{\mu\nu}+8\pi G\,^{(4)}T_{\mu\nu}^{(IM)},
\end{equation}
\begin{equation}\label{a14}
^{(4)}\Box \phi +V'_{ind}(\phi)=0,
\end{equation}
where $(')$ is denoting derivative with respect to the field $\phi$ and
\begin{equation}\label{dt1}
^{(4)}T^{(IM)}_{\mu\nu}=\frac{3}{8\pi G}\left[\left(\frac{d^{2}A}{d\psi^{2}}\right)_{\psi=\psi 
_0}+2\left(\frac{dA}{d\psi}\right)_{\psi=\psi _0}^{2}\right]e^{2A(\psi _0)}\,h_{\mu\nu}
\end{equation}
is the induced matter energy-momentum tensor on the geometrical background (\ref{a5}).

\section{ quantum cosmological confinement of the 5D scalar modes}

As an illustration of the previous formalism we shall study the cosmological case. In order to describe a 5D space-time with a 3D 
spatial expansion which in addition is spatially flat, we consider the line element
\begin{equation}\label{b1}
dS^{2}=e^{2A(\psi)}\left[dt^{2}-a^{2}(t)dr^{2}\right]-d\psi^{2},
\end{equation}
where $a(t)$ is the scale factor, $t$ is the cosmic time and $dr^{2}=dx^{2}+dy^{2}+dz^{2}$ being $\lbrace x,y,z\rbrace$ the usual 
cartesian coordinates. The 5D equation of motion (\ref{a6}) now reads
\begin{equation}\label{b2}
\ddot{\varphi}+3H(t)\dot{\varphi}-\frac{1}{a(t)}\nabla _{r}^{2}\varphi 
-e^{2A(\psi)}\left[4\frac{dA(\psi)}{d\psi}\frac{\partial\varphi}{\partial\psi}+\frac{\partial^{2}\varphi}{\partial\psi^{2}}\right]
=0,
\end{equation}
being $H(t)=\dot{a}/a$ the Hubble parameter and the dot denoting time derivative. Following a canonical quantization process
we impose the equal time commutation relations
\begin{eqnarray}\label{b6}
&&\left[\hat{\varphi}(t,\vec{r},\psi),\hat{\Pi}_{\varphi}^{t}(t,\vec{r'},\psi 
')\right]=i\frac{1}{a^{3}}\,e^{-2A(\psi)}\,\delta^{(3)}(\vec{r}-\vec{r'})\delta (\psi-\psi '),\\
\label{b7}
&& \left[\hat{\varphi}(t,\vec{r},\psi),\hat{\varphi}(t,\vec{r'},\psi ')\right]=\left[\hat{\Pi}_{\varphi}^{t}(t,\vec{r},\psi 
),\hat{\Pi}_{\varphi}^{t}(t,\vec{r'},\psi ')\right]=0,
\end{eqnarray}
where $\Pi _{\varphi}^{t}=\frac{\partial \, ^{(5)}L}{\partial \dot{\varphi}}= \sqrt{|\, ^{(5)}g|}\,\,g^{tt}\dot{\varphi} = 
e^{2A(\psi)}a^{3}\dot{\varphi}$ is the momentum conjugated to $\varphi$.

The operator $\hat{\varphi}$ is decomposed in Fourier modes
\begin{equation}\label{b7}
\hat{\varphi}(t,\vec{r},\psi)=\frac{1}{(2\pi)^{3/2}}\int d^{3}k_{r}dk_{\psi}\left[\hat{a}_{k_{r}k_{\psi}}\xi 
_{k_{r}k_{\psi}}(t,\psi)e^{i\vec{k}_{r}\cdot\vec{r}}+\hat{a}_{k_{r}k_{\psi}}^{\dagger}\xi 
_{k_{r}k_{\psi}}^{*}(t,\psi)e^{-i\vec{k}_{r}\cdot\vec{r}}\right],
\end{equation}
where the annihilation and creation operators $\hat{a}_{k_{r}k_{\psi}}$ and $\hat{a}_{k_{r}k_{\psi}}^{\dagger}$ satisfy the 
canonical commutation algebra
\begin{eqnarray}\label{b8}
&&\left[\hat{a}_{k_{r}k_{\psi}},\hat{a}_{k'_{r}k'_{\psi }}^{\dagger}\right]=\delta 
^{(3)}(\vec{k}_{r}-\vec{k'}_{r})\delta(k_{\psi}-k'_\psi),\\
\label{b9}
&& 
\left[\hat{a}_{k_{r}k_{\psi}},\hat{a}_{k'_{r}k'_{\psi}}\right]=\left[\hat{a}_{k_{r}k_{\psi}}^{\dagger},\hat{a}_{k'_{r}k'_{\psi}}^{
\dagger}\right]=0,
\end{eqnarray}
while the Wronskian condition for the $k_{r}k_{\psi}$-modes yields
\begin{equation}\label{b10}
\xi _{k_{r}k_{\psi}}\dot{\xi}_{k_{r}k_{\psi}}^{*}-\dot{\xi}_{k_{r}k_{\psi}}\xi 
_{k_{r}k_{\psi}}^{*}=i\frac{1}{a^{3}}\,e^{-2A(\psi)},
\end{equation}
with  the asterisk $(*)$ denoting complex conjugate.
Inserting (\ref{b7}) in (\ref{b2}) we obtain
\begin{equation}\label{b11}
\ddot{\xi}_{k_{r}k_{\psi}}+ 3H\dot{\xi} _{k_{r}k_{\psi}}+\left[a^{-2}k_{r}^{2}-e^{2A(\psi)}\left(4\overset{\star}{A}\frac{\partial 
}{\partial \psi}+\frac{\partial ^{2}}{\partial \psi ^{2}}\right)\right]\xi _{k_{r}k_{\psi}}=0
\end{equation}
which is the dynamical equation for the modes $\xi _{k_{r}k_{\psi}}(t,\psi)$.
By introducing the auxiliary field
\begin{equation}\label{b3}
\zeta _{k_{r}k_{\psi}}(t,\psi)=\exp\left[\frac{3}{2}\int H(t)dt\right]\exp\left[2A(\psi)\right]\xi _{k_{r}k_{\psi}}(t,\psi),
\end{equation}
we can write equation (\ref{b11}) as
\begin{equation}\label{b4}
\ddot{\zeta}_{k_{r}k_{\psi}}+\left[a^{-2}k_{r}^{2}-\frac{9}{4}H^{2}-\frac{3}{2}\dot{H}-e^{2A(\psi)}\left(4\overset{\star}{A}^{2} + 
2\overset{\star\star}{A} + \frac{\partial ^{2}}{\partial \psi ^2}\right)\right]\zeta _{k_{r}k_{\psi}}=0.
\end{equation}
Assuming that the $k_{r}k_{\psi}$-modes can be separated in the form $\zeta _{k_{r}k_{\psi}}(t,\psi)=\zeta _{k_{r}}(t)\zeta 
_{k_{\psi}}(\psi)$, the equation (\ref{b11}) yields
\begin{eqnarray}\label{b12}
\ddot{\zeta}_{k_{r}}+\left[a^{-2}k_{r}^{2}-\frac{9}{4}H^{2}-\frac{3}{2}\dot{H}-\beta^{2}\right]\zeta _{k_r}&=&0,\\
\label{b13}
\overset{\star\star}{\zeta}_{k_\psi}+\left[4\overset{\star}{A}^{2}+2\overset{\star\star}{A}-\beta^{2}e^{-2A(\psi)}\right]\zeta 
_{k_\psi}&=& 0,
\end{eqnarray}
being $\beta$ a separation constant. The Wronskian condition (\ref{b10}) now becomes
\begin{equation}\label{b14}
\zeta _{k_{r}}\dot{\zeta} _{k_r}^{*}-\dot{\zeta}_{k_r}\zeta _{k_r}^{*}=i,\qquad \zeta _{k_\psi}\zeta _{k_\psi}^{*}=e^{2A(\psi)}.
\end{equation}
Expression (\ref{b13}) means that the stability of the $k_{\psi}$-modes depends strongly of the warping factor. In other words, 
the mode dynamics along the fifth dimension is driven by the warping factor. Moreover, as we saw in the previous section the 
warping factor not solely controls the dynamics of the modes in the $k_{\psi}$-direction, but also is a preponderant factor in the 
induction of the scalar potential $V(\phi)$ on $\Sigma _{0}$.\\

As an illustrative example we consider the warping factor $A(\psi)=ln(\psi/\psi  _0)$ which corresponds to the line element
\begin{equation}\label{b15}
dS^{2}=\left(\frac{\psi}{\psi _0}\right)^{2}\left[dt^{2}-a^{2}(t)dr^{2}\right]-d\psi^{2},
\end{equation}
which is a generalization of the Ponce de Leon metric \cite{PLM} on which $a(t)=\exp(2t/\psi _0)$.
Thus, equation (\ref{b13}) reduces to
\begin{equation}\label{b16}
\overset{\star\star}{\zeta}_{k_\psi}+\left(\frac{2-\beta^{2}\psi _{0}^{2}}{\psi^{2}}\right)\zeta _{k_\psi}=0
\end{equation}
whose general solution is given by
\begin{equation}\label{b17}
\zeta _{k_{\psi}}(\psi)=B_{1}(k_{\psi})\psi ^{\frac{1}{2}(1+\sqrt{4\beta^{2}\psi 
_{0}^{2}-7})}+B_{2}(k_{\psi})\psi^{\frac{1}{2}(1-\sqrt{4\beta^{2}\psi _{0}^{2}-7})}.
\end{equation}
Considering for simplicity a de-Sitter expansion $a(t)=a_{0}e^{H_{0}t}$, the equation of motion for the $k_{r}$-modes (\ref{b12}) 
now reads
\begin{equation}\label{b18}
\ddot{\zeta}_{k_r}+\left[k_{r}^{2}a_{0}^{-2}e^{-2H_{0}t}-\frac{9}{4}H_{0}^{2}-\beta^{2}\right]\zeta _{k_{r}}=0.
\end{equation}
The general solution of (\ref{b18}) have the form
\begin{equation}\label{b19}
\zeta _{k_r}(t)=D_{1}(k_{r}){\cal H}_{\nu}^{(1)}[\frac{k_r}{a_{0}H_{0}}e^{-H_{0}t}]+D_{2}(k_{r}){\cal 
H}_{\nu}^{(2)}[\frac{k_r}{a_{0}H_{0}}e^{-H_{0}t}]
\end{equation}
where $\nu= 1/(2H_{0})\sqrt{9H_{0}^{2}+4\beta^{2}}$ and ${\cal H}_{\nu}^{(1,2)}$ are the first and second kind Hankel functions. 
Using the normalization condition (\ref{b14}), which now is $\xi _{k_{\psi}}\xi _{k_{\psi}}^{*}=(\psi/\psi _0)^{2}$, and selecting 
the Bunch-Davies vacuum \cite{BDavies} for de-Sitter space $B_{2}(k_{\psi})=0$, $D_{1}(k_{r})=0$ we obtain that the unique 
normalizable  $k_{\psi}$-mode is obtained by setting $\beta =\pm\sqrt{2}/\psi _0$. In this case the normalized solution for the 
$k_{r}k_{\psi}$-modes $\zeta _{k_{r}k_{\psi}}$ is
\begin{equation}\label{b20}
\bar{\zeta}_{k_r}(t,\psi)\equiv \zeta _{k_{r}k_{\psi}}(t,\psi)=\frac{i}{2}\sqrt{\frac{\pi}{H_{0}}}\left(\frac{\psi }{\psi 
_0}\right){\cal H}_{\nu}^{(2)}\left[\frac{k_{r}}{a_{0}H_{0}}e^{-H_{0}t}\right],
\end{equation}
and therefore
\begin{equation}\label{b21}
\bar{\xi}_{k_{r}}(t)\equiv\xi _{k_{r}k_{\psi}}(t,\psi)=\frac{i}{2}\sqrt{\frac{\pi}{H_{0}}}e^{-\frac{3}{2}H_{0}t}{\cal 
H}_{\nu}^{(2)}\left[\frac{k_{r}}{a_{0}H_{0}}e^{-H_{0}t}\right],
\end{equation}
with $\nu =[1/(2H_{0})]\sqrt{9H_{0}^{2}+8\psi _{0}^{-2}}$. Clearly, the modes $\bar{\xi} _{k_{r}}$ do not  exhibit any dependence 
of the fifth coordinate. This fact can be interpreted as a kind of confinement of the quantum scalar modes $\bar{\xi}_{k_{r}}(t)$ 
on the hypersurfaces $\Sigma _{0}:\psi=\psi _{0}$, since the modes of $\varphi$ do not propagate along the fifth dimension. Note 
that  this confinement is obtained naturally from the theory without the introduction of a confining scalar potential.

\section{4D induced inflation}

Now we are able of treating the 4D inflationary case derived from the induced 4D action (\ref{a13}). According to the formalism 
exposed in section II, the induced 4D line element is given by
\begin{equation}\label{c1}
ds^{2}=h_{\mu\nu}(x)dx^{\mu}dx^{\nu}=dt^{2}-a^{2}(t)dr^{2}.
\end{equation}
According to equations (\ref{pr1}) and (\ref{dt1}) the induced matter density and pressure can be separated as
\begin{equation}\label{dt2}
\rho _{eff}=\rho _{\phi}+\rho _{(IM)},\qquad P_{eff}=P_{\phi}+P_{(IM)},
\end{equation}
where
\begin{eqnarray}\label{dt3}
\rho _{\phi}&=&\frac{1}{2}\dot{\phi}^{2}+\frac{1}{2a^2}(\nabla _{r}\phi)^{2}+V_{ind}(\phi),\\
P _{\phi}&=&\frac{1}{2}\dot{\phi}^{2}+\frac{1}{2a^2}(\nabla _{r}\phi)^{2}-V_{ind}(\phi),\\
\rho _{(IM)}&=&\frac{3}{8\pi G}\left[\left(\frac{d^{2}A}{d\psi ^2}\right)_{\psi=\psi _0}+2\left(\frac{dA}{d\psi}\right)_{\psi=\psi 
_0}^{2}\right]=-P_{(IM)}.
\end{eqnarray}
This way we have two contributions one due to the scalar field and another due to the assumed action of the fifth dimension of 
inducing matter on the 4D space-time \cite{Wes}. the effective equation of state for a slow-roll conditions on the inflaton field 
$\phi$, can be written as $P_{eff}\simeq -\rho _{eff}$.\\

In order to determine an exact form of the induced scalar potential (\ref{a11}), we consider for simplicity the same warping 
factor $A(\psi)=ln(\psi/\psi _0)$. Under this considerations the equation (\ref{a8}) has the general solution
\begin{equation}\label{c2}
Q(\psi)=C_{1}\psi ^{(1/2)[-3+\sqrt{9-4\alpha\psi _{0}^{2}}\,]}+C_{2}\psi ^{(-1/2)[3+\sqrt{9-4\alpha\psi _{0}^{2}}\,]}
\end{equation}
Choosing $C_{2}=0$, the 4D induced scalar potential (\ref{a11}) reads
\begin{equation}\label{c3}
V_{ind}(\phi)=\frac{1}{2}\left[\frac{\sqrt{9-4\alpha\psi _{0}^{2}}-3}{2\psi _0}\right]^{2}\phi^{2}(t,\vec{r}).
\end{equation}
The field equation for the scalar field $\phi(t,\vec{r})$ is then
\begin{equation}\label{c4}
\ddot{\phi}+3H(t)\dot{\phi}-a^{-2}(t)\nabla _{r}^{2}\phi + \left(\frac{\sqrt{9-4\alpha\psi _{0}^{2}}-3}{2\psi 
_0}\right)^{2}\phi=0.
\end{equation}
The field operators $\hat{\phi}(t,\vec{r})$ and $\hat{\Pi}_{\phi}^{t}=\partial \,^{(4)}L/(\partial \dot{\phi})=a^{3}\dot{\phi}$ 
satisfy the equal time algebra
\begin{equation}\label{c5}
\left[\hat{\phi}(t,\vec{r}),\hat{\Pi}_{\phi}^{t}(t,\vec{r'})\right]=i\frac{1}{a^{3}}\delta ^{(3)}(\vec{r}-\vec{r'}),\quad 
\left[\hat{\phi}(t,\vec{r}),\hat{\phi}(t,\vec{r'})\right]=\left[\hat{\Pi}_{\phi}^{t}(t,\vec{r}),\hat{\Pi}_{\phi}^{t}(t,\vec{r'}) 
\right]=0.
\end{equation}
Thus, the field operator $\hat{\phi}(t,\vec{r})$ can be decomposed in Fourier modes as
\begin{equation}\label{c6}
\hat{\phi}(t,\vec{r})=\frac{1}{(2\pi)^{3/2}}\int 
d^{3}k_{r}\left[\hat{a}_{k_{r}}h_{k_{r}}(t)e^{i\vec{k}_{r}\cdot\vec{r}}+\hat{a}_{k_r}^{\dagger}h_{k_r}^{*}(t)e^{-i\vec{k}_{r}\cdot
\vec{r}}\right],
\end{equation}
where the annihilation and creation operators $\hat{a}_{k_r}$ and $\hat{a}_{k_r}^{\dagger}$ satisfy the commutation relations
\begin{equation}\label{c7}
\left[\hat{a}_{k_{r}},\hat{a}_{k'_{r}}^{\dagger}\right]=\delta ^{(3)}(\vec{r}-\vec{r'}),\quad 
\left[\hat{a}_{k_r},\hat{a}_{k'_r}\right]=\left[\hat{a}_{k_r}^{\dagger},\hat{a}_{k'_{r}}^{\dagger}\right]=0.
\end{equation}
The Wronskian condition for the $k_{r}$-modes $h_{k_r}$ is given by
\begin{equation}\label{c8}
h_{k_r}\dot{h}_{k_r}^{*}-\dot{h}_{k_r}h_{k_r}^{*}=\frac{i}{a^3}.
\end{equation}

Now, implementing the transformation $\phi(t,\vec{r})=\exp [(-3/2)\int H(t)\,dt]\,\chi (t,\vec{r})$, the expression (\ref{c4}) 
becomes
\begin{equation}\label{c10}
\ddot{\chi}-a^{-2}\nabla _{r}^{2}\chi -\left[\frac{9}{4}H^{2}+\frac{3}{2}\dot{H}-m^{2}\right]\chi =0,
\end{equation}
where $m^{2}=\left[(\sqrt{9-4\alpha\psi _{0}^{2}}-3)/(2\psi _0)\right]^{2}$ plays the role of a geometrical inflaton ``mass". 
Hence, the next Fourier expansion is valid
\begin{equation}\label{c11}
\chi (t,\vec{r})=\frac{1}{(2\pi)^{3/2}}\int 
d^{3}k_{r}\left[\hat{a}_{k_{r}}u_{k_{r}}(t)e^{i\vec{k}_{r}\cdot\vec{r}}+\hat{a}_{k_{r}}^{\dagger}u_{k_r}^{*}(t)e^{-i\vec{k}_{r} 
\cdot\vec{r}}\right],
\end{equation}
being now $u_{k_r}(t)=\exp [(3/2)\int H(t)dt]\,h_{k_r}(t)$ and $u_{k_r}^{*}(t)=\exp [(3/2)\int H(t)dt]\,h_{k_r}^{*}(t)$. In terms 
of the new $k_{r}$-modes $u_{k_r}$, the Wronskian condition (\ref{c8}) reads
\begin{equation}\label{c12}
u_{k_r}\dot{u}_{k_r}^{*}-\dot{u}_{k_r}u_{k_r}^{*}=i.
\end{equation}
Considering for simplicity a de Sitter expansion $a(t)=a_{0}\exp[H_{0}t]$, the evolution equation for the $k_{r}$-modes $u_{k_r}$ 
is
\begin{equation}\label{c13}
\ddot{u}_{k_r}+\left[k_{r}^{2}a_{0}^{-2}e^{-2H_{0}t}-\frac{9}{4}H_{0}^{2}-m^{2}\right]u_{k_r}=0.
\end{equation}
Stable solutions of this equation can be obtained for $k_{r}> k_{0}$, where $k_{0}(t)=[(9/4)H_{0}^{2}+m^{2}]^{1/2}\,\exp[H_{0}t]$. 
The general solution of (\ref{c13}) is given by
\begin{equation}\label{c14}
u_{k_r}(t)=F_{1}{\cal H}_{\nu _s}^{(1)}[Z(t)] + F_{2}{\cal H}_{\nu _s}^{(2)}[Z(t)],
\end{equation}
with $\nu _s =[1/(2H_{0})]\sqrt{9H_{0}^{2}+4m^{2}}$, $Z(t)=[k_{r}/(a_{0}H_{0})\,e^{-H_{0}t}]$, and $F_{1}$, $F_{2}$ being 
integration constants. Selecting the Bunch-Davies vacuum $F_{1}=0$ the usual normalization is achieved by using (\ref{c12}). 
Therefore a normalized solution is
\begin{equation}\label{c15}
u_{k_r}(t)=\frac{i}{2}\sqrt{\frac{\pi}{H_0}}{\cal H}_{\nu _s}^{(2)}[Z(t)].
\end{equation}
The squared quantum fluctuations on the IR-sector $(k_{r}\ll k_{0})$ are given by
\begin{equation}\label{c16}
\left<\phi ^{2}(t)\right> _{IR}=\frac{e^{-3H_{0}t}}{2\pi ^{2}}\int _{0}^{\upsilon 
k_{H}}\frac{dk_{r}}{k_{r}}k_{r}^{3}\left.[u_{k_r}(t)u_{k_{r}}^{*}(t)]\right|_{IR}
\end{equation}
where $\upsilon =k_{max}^{IR}/k_{p}\ll 1$ is a dimensionless parameter, being $k_{max}^{IR}=k_{H}(t_i)=k_{0}(t_{i})$ the wave 
number related to the Hubble radius at the time $t_{i}$ when the modes re-enter to the horizon. Besides $k_{p}$ is the Planckian 
wave number. For a Hubble parameter value $H_{0}=0.5\times 10^{-9}\,\, M_{p}$, values of $\upsilon$ within the interval $10^{-5}$ 
to $10^{-8}$ corresponds  to a number of e-foldings $N_{e}=63$ \cite{GEM}. Thus, considering the asymptotic expansion for the 
Hankel function ${\cal H}_{\nu _s}^{(2)}[Z(t)]\simeq(-i/\pi)\Gamma (\nu _{s})(Z/2)^{-\nu _s}$, the equation (\ref{c16}) becomes
\begin{equation}\label{c17}
\left<\phi ^{2}\right>=\frac{2^{2\nu _{s}}}{8\pi}\left.\frac{a_{0}^{3}H_{0}^{2}}{3-2\nu _{s}}\Gamma ^{2}(\nu 
_{s})\left(\frac{k_{r}}{a(t)H_{0}}\right)^{3-2\nu _{s}}\right|_{\upsilon k_{H}}\,.
\end{equation}
and the corresponding power spectrum ${\cal P}_{\phi}(k_{r})\sim [k_{r}/(aH_{0})]^{3-2\nu _s}$. Clearly, this spectrum becomes 
nearly scale invariant when $m\ll 1$. Furthermore, the spectral index is given by $n_{s}=4-2\nu 
_{s}=4-3\sqrt{1+[2m/(3H_{0})]^{2}}$. Hence, knowing from the observational data that \cite{ns} $0.94\leq n_{s}\leq 1$, we can 
establish that $0\leq m\leq 0.3 H_{0}$. This interval corresponds to $-1/4(H_{0}/\psi _{0})[0.36H_{0}\psi _{0}+3.6]\leq \alpha\leq 
0$. Therefore, for every hypersurface $\psi =\psi _0$ the value of $\alpha$ can change in such a way that $m$ remains taking 
values in the same interval.

\section{Final comments and conclusions}

We have studied a scalar field formalism from a pure kinetic 5D scalar field $\varphi$ on the class of 5D warped product spaces. 
Within this new approach we have developed a consistent manner to induce a 4D lagrangian density for a true 4D scalar field 
$\phi$, from a 5D lagrangian density that describes the kinetic scalar field $\varphi$. As a first application of the formalism we 
have studied inflationary cosmology. This new approach enables us to have a 4D inflationary formalism where the scalar potential 
$V(\phi)$ is induced due to the motion of the field $\varphi$ with respect to the fifth coordinate $\psi$, in an special manner. 
The 4D formalism is constructed on a family of hypersurfaces given by the constant foliation $\Sigma _{0}:\psi =\psi _{0}$. 
Something that is very important to stress is that the formalism is based strongly on the separability of the field $\varphi$ and 
on the properties of the warped product spaces. Indeed, the applicability of the general mechanism of induction of 4D potentials 
could go beyond cosmology. This is because the minimal conditions in order to the mechanism works are the separability of the 5D 
scalar field $\varphi$ and the separability of the components of the metric $g_{\mu\nu}(x,\psi)$ and $g_{\psi\psi}(x,\psi)$. Which 
of course the class of warped product spaces fits in. In the case of a space-time dependence in $g_{\psi\psi}$ the induced 4D 
potential will be of the form $V[x,\phi(x)]=(1/2)M^{2}(x)\phi^{2}(x)$. In other words, the induced potential not only will depend 
on the true scalar field $\phi$, but also on the space-time coordinates. This way the mechanism could explain in a geometrical 
manner with an extra dimension the appearance of local scalar potentials which have an extra space-time dependence in contexts 
that are not necessarily cosmological. In an inflationary scenario we have obtained that at least for a class of warping factors  
the quantum modes associated to the 5D scalar field $\varphi$ exhibit a natural quantum confinement to the 4D hypersurfaces 
$\Sigma _{0}$. This characteristic is merely quantum in origin since it is derived as a consequence of the employment of the 
quantum commutation relations in 5D. An interesting implication of this quantum behavior is that for instance for a warping factor 
$A(\psi)=ln(\psi/\psi _{0})$, if we choose $\psi _{0}=\sqrt{3/\Lambda}$ with $\Lambda$ the cosmological constant, 5D classical 
particles have not confinement to hypersurfaces solely due to gravitational effects \cite{CRomero}. However, in the case of a 5D 
quantum scalar field confinement of the quantum modes is achieved naturally. Finally, as it is well-known in many higher 
dimensional cosmological theories, one issue consists in to explain the fact that the observable universe seems to be confined to 
a particular hypersurface. In our formalism the part of the 4D dynamics that in principle could depend of the value of $\psi 
_{0}$, is the part that contains the information of the 4D induced potential $V(\phi)=(1/2)m^{2}\phi^{2}(x)$. However, the 
presence of a separation constant in (\ref{a8}) makes possible that $m$, as it is  defined from (\ref{a11}), can remain invariant 
for any value of $\psi _{0}$. One example of this fact is the final result of the de-Sitter inflationary model in this letter, 
where the interval $0\leq m\leq 0.3 H_{0}$ is maintained invariant for $-1/4(H_{0}/\psi _{0})[0.36H_{0}\psi _{0}+3.6]\leq 
\alpha\leq 0$, independently of the value of $\psi _0$.

\section*{Acknowledgements}
\noindent
JEMA acknowledges CNPq-CLAF and UFPB (Brazil) for financial
support. \\

\end{document}